\documentclass[aps,pre,twocolumn,showpacs,superscriptaddress]{revtex4}
\usepackage{amsmath}
\usepackage{amssymb}
\usepackage{color}
\usepackage{graphicx}
\begin{document}
\title{Solvent-Dependent Critical Properties of Polymer Adsorption}
\author{Jo\~ao~A.\ Plascak}
\email{E-mail: pla@physast.uga.edu}
\affiliation{Departamento de F\'isica, 
Universidade Federal da Para\'iba, 
58051-970 Jo\~ao  Pessoa, PB - Brazil}
\affiliation{Departamento de F\'{\i}sica,
Universidade Federal de Minas Gerais, 31270-901 Belo Horizonte, MG -
Brazil}
\affiliation{Center for Simulational Physics, The University of Georgia, 
Athens, GA 30602, USA}
\author{Paulo~H.~L.\ Martins}
\email{E-mail: pmartins@fisica.ufmt.br}
\affiliation{Center for Simulational Physics, The University of Georgia, 
Athens, GA 30602, USA}
\affiliation{Instituto de F\'isica, Universidade Federal de Mato Grosso,
78060-900 Cuiab\'a, MT - Brazil}
\author{Michael Bachmann}
\email[E-mail: ]{bachmann@smsyslab.org}
\homepage[\\ Homepage: ]{http://www.smsyslab.org}
\affiliation{Departamento de F\'{\i}sica,
Universidade Federal de Minas Gerais, 31270-901 Belo Horizonte, MG -
Brazil}
\affiliation{Center for Simulational Physics, The University of Georgia, 
Athens, GA 30602, USA}
\affiliation{Instituto de F\'isica, Universidade Federal de Mato Grosso,
78060-900 Cuiab\'a, MT - Brazil}
\date{\today}
\begin{abstract}
Advanced chain-growth computer simulation methodologies have been employed for 
a systematic statistical analysis of the critical behavior of a polymer 
adsorbing at a substrate. We use finite-size scaling techniques to 
investigate the solvent-quality dependence of critical exponents, critical 
temperature, and the structure of the phase diagram. Our study covers all 
solvent effects from the limit of \emph{super-self-avoiding walks},
characterized by 
effective monomer-monomer repulsion, to poor solvent conditions that enable the 
formation of compact polymer structures. The results significantly benefit from 
taking into account corrections to scaling.
\end{abstract}
\maketitle
The study of polymer adsorption on a flat solid
surface has been extensively investigated for more than 60 years~\cite{sinha}.
Understanding generic properties of this process is not only relevant for 
potential technological and 
biological applications~\cite{gennes,milner,meredith,diaz,bach14}, but also
for more basic insights into phenomena such as adhesion, surface coating, 
wetting, and
adsorption chromatography~\cite{fleer}. 
In dilute solution, polymers are
independent of each other and surface
effects affect the structure formation process individually. 
Conformational properties are thus basically influenced by
heat-bath temperature, solvent quality, and the strengths of
monomer-monomer
and monomer-surface interactions. In general, at sufficiently high 
temperatures and
good solvent conditions, the polymer chain favors a disordered random 
(typically expanded) geometric structure and it is, for the gain of 
translational entropy, desorbed.
However, below a certain threshold temperature, an attractive 
interaction with the surface can energetically overcompensate the entropic 
freedom of the 
chain and  
chain segments get adsorbed at the surface. In consequence, 
a continuous
adsorption-desorption (A-D) transition~\cite{eisen} occurs at a critical 
temperature $T_a$,
separating the desorbed phase, which is dominant for $T>T_a$, from the 
phase governed by adsorbed polymer conformations for $T<T_a$.

An appropriate order parameter for this A-D transition is
$n_s= N_s/N$,
where $N_s$ is the number of monomers in contact with the surface and
$N$ is the total length of the chain. In discrete representation, a 
monomer is in contact with the substrate if a monomer and a substrate 
bead 
are nearest neighbors on the lattice.
In the desorbed phase
($T>T_a$), $n_s\rightarrow 0$ for very long chains 
($N\to\infty$). The power laws
$\langle N_s \rangle\sim N^\phi\quad\mbox{or}\quad\langle n_s \rangle\sim
N^{\phi-1}$,
where $\phi$ is a crossover exponent~\cite{eisen},
are expected to hold at the transition temperature $T_a$.

In three dimensions, the consistent estimation of a precise value of 
the crossover
exponent is a longstanding and still open problem. Various 
values around $\phi\approx 0.5$ have been 
proposed~\cite{eisen,meiro,hegger,metz1,metz2,descas,grass,binder,luo,%
taylor} (including the long-term conjecture of $\phi=0.5$ being
super-universal and independent of dimension \cite{hegger}), but 
the posted uncertainties are much smaller than the deviations among the 
estimates. This indicates that there might be a systematic issue which has not 
yet been properly addressed. The numerical value 
of $\phi$  
depends strongly on the precise estimate of the critical temperature
$T_a$.

In most
previous studies only good solvent conditions were considered, i.e.,
the intrinsic interaction between nonbonded monomers has 
been widely neglected. However, it is 
also important to understand how the
scaling behavior
depends on the solvent conditions and their influence on the 
transition properties as represented in the 
phase diagram, parametrized by temperature and solvent quality.

In this Letter, we systematically study the solvent dependence of 
critical properties of the A-D 
transition 
of linear, flexible polymer chains grafted to a substrate. Our results aim at 
providing the quantitative foundation for the understanding of the critical 
adsorption behavior of entire classes of hybrid polymer-substrate systems. 
For this purpose, we utilize the
similarity of the A-D transition with phase transitions in magnetic 
systems~\cite{luo,puli}, and employ finite-size scaling 
theory for the characterization of the critical properties. 
Corrections-to-scaling effects are considered as well to 
take into account the finite length of the simulated polymers chains.

The polymer model consists of $N$
identical beads occupying sites on a three-dimensional (simple-cubic) 
lattice. The polymer chain represents an interacting self-avoiding walk with 
short-range interactions between pairs of nonbonded monomers and monomers and 
substrate sites. Solvent conditions are changed by
varying the energy scales of these competing interactions.

Adjacent monomers
in the polymer chain have unity bond length. We
consider a grafted polymer with one end covalently, and permanently, bound to
the surface.
Each pair of nearest-neighbor nonbonded monomers possesses an energy
$-\epsilon_m$. Thus, the key parameter for the energetic state of the
polymer itself is the number of monomer-monomer contacts, $N_m$. The flat
homogeneous and impenetrable substrate is located in the $z=0$ plane, and 
monomer locations are restricted to $z>0$. All monomers lying in the $z=1$ 
plane are 
considered to be in contact with the substrate, and an energy $-\epsilon_s$ is
attributed to each one of these surface contacts. Hence, the energetic
contribution due to the interaction with the substrate is given by the number of
surface contacts of the polymer, $N_s$.

The total energy of the model can be written as
\begin{equation}
E_s(N_s, N_m) =-\epsilon_s N_s -\epsilon_m N_m= -\epsilon_s(N_s + s N_m), 
\label{en}
\end{equation}
where $s = \epsilon_m/\epsilon_s$ is the ratio of the respective
monomer-monomer and monomer-substrate energies.  
Actually, $s$ controls the solvent quality in such a way that 
larger $s$ values favor the formation of monomer-monomer contacts 
(poor solvent), whereas smaller values lead to a stronger binding to the 
substrate. 
For convenience, we set $\epsilon_s = 1$ meaning that all energies are 
measured in units of the monomer-substrate interaction.

For the simulation of the model, we used the contact-density chain-growth 
algorithm~\cite{bach14}, which extends earlier chain-growth 
methods~\cite{rosenbluth1,grass1,hsu1,bj1,prellberg1}. Consequently,
the 
contact density (or number of states) $g(N_s, N_m)$ is directly obtained in the
simulation for any possible 
pair
$N_s$ and $N_m$. It is independent of temperature and the ratio of the 
interactions $s$. Thus, the temperature $T$ and 
the
solubility parameter $s$ are external parameters that can be set after the
simulation is finished.
We generated $10^8-10^9$ chains with lengths $N = 16$, $32$, $64$, $128$, 
$256$, $400$, and $503$ monomers.

The contact density $g(N_s, N_m)$ is a versatile quantity in that all relevant 
energetic thermodynamic observables
can be obtained by simple reweighting. For
instance, for a given pair $N_s$ and $N_m$, the restricted
partition function $Z_{T,s}^r (N_s, N_m)$ can be defined as
$Z_{T,s}^r (N_s, N_m) = g(N_s, N_m) \exp[(N_s + s N_m)/k_\mathrm{B} T ],$
from which the canonical partition function is obtained as
$Z_{T,s} = \sum_{N_s,N_m}Z_{T,s}^r (N_s, N_m).$
Similarly, the mean value of any quantity $Q(N_s, N_m)$ can also be computed by 
reweighting,
\begin{equation}
\langle Q \rangle=\frac{1}{Z_{T,s}} \sum_{N_s,N_m} Q(N_s, N_m) g(N_s, N_m)
e^{(N_s + s N_m )/k_\mathrm{B} T}.
\end{equation}
In the following, we set $k_B=1$. Apparently, contact entropy, free energy, 
average number of surface
contacts $N_s$, average number of monomer-monomer contacts $N_m$, heat
capacity, cumulants, etc.\ are examples of functions that are easily calculable
for any values of $T$ and $s$ once $g(N_s, N_m)$ is known.

The scaling properties of generic energetic quantities, such as maxima 
of specific-heat curves~\cite{bach05,kraw}, 
have proven to be rather unsuitable for a systematic scaling 
analysis~\cite{janse}, whereas the scaling behavior of the partition function
turned out to be more insightful~\cite{grass,binder}.
We investigate the scaling properties of the order parameter 
and its derivatives similarly to Ref.~\cite{luo}. 
However, going beyond the standard approach, we take into account 
corrections to scaling and
use for our analysis convenient temperature derivatives of the order parameter, 
as well as
scaling properties of the A-D transition temperature and the fourth-order
cumulant of the order parameter.

From the simulation results we estimate $\langle n_s\rangle$, the
fourth-order Binder
cumulant
\begin{equation}
U_{4}(T)=1-\frac{\left<n_s^{4}\right>}{3\left<n_s^{2}\right>^{2}},
\label{u4}
\end{equation}
and the logarithmic temperature derivative
\begin{equation}
\Gamma_{n_s}=\frac{d\ln\langle n_s\rangle}{dT}
\label{dlns}
\end{equation}
for each polymer length $N$.
It is well known that, according to finite-size scaling (FSS) theory for
second-order phase transitions, the order
parameter $\langle n_s\rangle$ should scale close to the critical temperature as
\begin{equation}
\langle n_s\rangle=N^{\phi-1}f_{n_s}(x)\left[1+A_{n_s}(x)N^{-\omega}\right],
\label{nscal}
\end{equation}
where corrections to scaling due to the finite polymer
length have been taken into account. The corresponding fourth-order cumulant of 
the order parameter 
$U_4$ given by Eq.~(\ref{u4}) should be 
independent of the chain length $N$ for very long chains~\cite{bincum}, 
and the maximum value of $\Gamma_{n_s}$, given by Eq.~(\ref{dlns}), supposedly 
scales like
\begin{equation}
\Gamma_{n_s}^\mathrm{max}=N^{1/\delta}f_d(x) 
\left[1+A_d(x)N^{-\omega} \right].
\label{dscal}
\end{equation}
In these equations, $\phi$ is the crossover exponent as defined in
Ref.~\cite{binder},  $\delta$  is the equivalent of the critical exponent
of the correlation length, $\nu$, in ordinary magnetic continuous phase
transitions, and $f_{n_s}(x)$ and $f_d(x)$ are FSS functions 
with 
$x=(T-T_{a})N^{1/\delta}$ being the scaling variable. 
The second term in the brackets in Eqs.~(\ref{nscal}) and~(\ref{dscal}) 
approximates all corrections to scaling by a single term, 
where $\omega$ is the leading correction-to-scaling exponent 
and $A_{n_s}(x)$ and $A_d(x)$ are non-universal functions
(see, for instance, Ref.~\cite{landaupa}).
\begin{figure}
\centerline{\includegraphics[width=8.5cm]{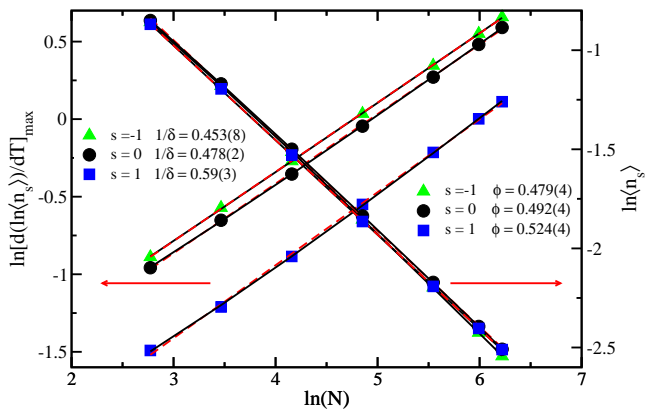}}
\caption{\label{fig:nuphi}(Color online) Logarithm of the maximum value of 
the 
order parameter derivative $\Gamma_{n_s}$,
defined in Eq.~(\ref{dlns}), and logarithm of the order parameter 
$\langle n_s\rangle$ as functions of the logarithm of the polymer
length $N$ for different $s$ values. The dots correspond to the
simulation results and the lines are the best fits according to  
Eqs.~(\ref{nscal}) and~(\ref{dscal}), 
without corrections to scaling (linear fit, assuming $A_d(x)=0$) and
with scaling corrections ($A_d(x)\ne 0$). The given numerical 
estimates include the corrections to scaling.}
\end{figure}

Accordingly, for the critical temperature the following scaling law holds, 
which is
also used in analogy to continuous transitions in magnetic models,
\begin{equation}
T_N= T_a + N^{1/\delta}f_{T}(x)\left[1+A_{T}(x)N^{-\omega}\right].
\label{tscal}
\end{equation}
Thus, the procedure we can follow is quite standard. From the 
simulations, we determine the exponent $1/\delta$ by using 
Eq.~(\ref{dscal}), which depends only on $\Gamma_{n_s}^\mathrm{max}$. In this 
case, we consider $f_d(x)$ and $A_d(x)$ as
constants (we do not expect them to vary appreciably since the maximum 
positions
should occur at temperatures close to the critical one). With this exponent at
hand, the critical temperature $T_a$ is obtained from Eq.~(\ref{tscal}) and
with it we estimate 
the crossover exponent
$\phi$ by using Eq.~(\ref{nscal}), in which case we can choose $x=0$.

As a test for the performance of the scaling 
approach for the data obtained in our simulations, let us first discuss
results for good solvent
conditions, $s=0$. In this case, we can compare with previously published 
results obtained with different methods.
Figure~\ref{fig:nuphi} shows the logarithm of the maximum value of the 
derivative
given in Eq.~(\ref{dlns}) as a function of the logarithm of the polymer length
$N$ for different solvent conditions, including the $s=0$ case for 
which the linear fit yields $1/\delta=0.448(3)$. Taking into account 
corrections to scaling we find $1/\delta=0.478(2)$, 
which indicates that corrections to scaling are relevant. Both estimates are, 
however, significantly smaller than the value reported in Ref.~\cite{luo},
$1/\delta=0.56$, which was obtained by a different approach.

\begin{figure}
\centerline{\includegraphics[width=8.5cm]{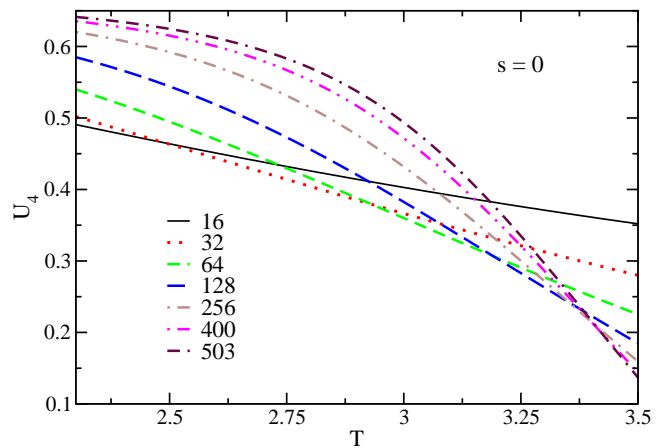}}
\caption{\label{fig:s0u4}(Color online) Fourth-order Binder cumulant $U_4$ as 
a function of
the temperature $T$ for different chain sizes for $s=0$.}
\end{figure}

\begin{figure}
\centerline{\includegraphics[width=8.5cm]{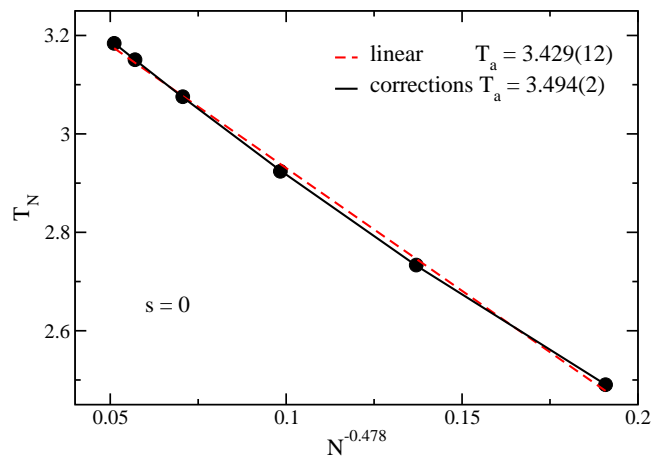}}
\caption{\label{fig:s0ta}(Color online) Transition temperature estimates $T_N$ 
as a function of
$N^{-1/\delta}$ for $s=0$. The dots correspond to the crossings of the
fourth-order Binder cumulant for chain lengths $N\ge32$ with the result for
$N=16$, as shown in Fig.~\ref{fig:s0u4}. 
The lines are the best fits according to Eq.~(\ref{tscal}), 
without corrections to scaling (linear fit, i.e., $A_T(x)=0$) and
with scaling corrections ($A_T(x)\ne 0$).}
\end{figure}

The fourth-order Binder cumulant, as a function of the temperature, is shown in
Fig.~\ref{fig:s0u4}.
One can clearly see that there is a systematic crossing of the curves for 
the
longer chains with $N\ge 32$ with the curve of the shortest, $N=16$. 
Considering these crossings
as finite-length estimates $T_N$ of the adsorption transition temperature, we 
plot the crossing points for $N\ge 32$ in Fig.~\ref{fig:s0ta}. For the $N$ 
dependence we make use of the ansatz (\ref{tscal}) with our previous 
estimate of the exponent $1/\delta\approx 0.478$. 

It is obvious that the inclusion of corrections to scaling is necessary in 
this case and our estimate 
$T_a=3.494(2)$ is very close to the most 
recently reported value  
$T_a=3.500(1)$ by Klushin et al.~\cite{binder}, who employed a 
different estimation method. 

After the critical temperature has been evaluated, we can utilize the
scaling relation~(\ref{nscal}) to determine the crossover exponent $\phi$. The 
results are
included in Fig.~\ref{fig:nuphi}. 
Although not visible in the scale used in the figure,
the corrections to scaling are important in this case, too. The thus computed 
value
$\phi=0.492(4)$ is also comparable with the
estimate given in 
Ref.~\cite{binder}, $\phi=0.483(2)$.

From the above results, we can conclude that the present approach and the data 
obtained from our simulations reproduce the scaling behavior for  
the special case of a non-interacting self-avoiding walk ($s=0$) 
very well. Results for the critical temperature of adsorption and the crossover 
exponent are in good agreement compared to the values previously
obtained by means of other procedures. 

Our method 
has the advantage that we can also analyze the structural behavior under
other 
solvent conditions for the polymer by varying the  
solvent parameter $s$ without the need of performing any additional
simulation.
The scaling behavior of the thermodynamic quantities for 
other $s$ values is qualitatively similar to the $s=0$ case presented
in 
Figs.~\ref{fig:nuphi}--\ref{fig:s0ta}, but the character of the adsorption 
transition changes. For poor solvent, i.e., $s>0$, desorbed and adsorbed 
polymer conformations are much more compact. The self-interacting polymer 
undergoes a collapse and additional freezing transition and both transitions 
compete with the adsorption transition, depending on the solvent conditions.
From the estimates for transition temperatures and critical exponents, we
find that the 
specific parametrization of the critical behavior depends on the solvent 
quality. As Fig.~\ref{fig:nuphi} shows, the values of the exponents 
obtained for 
$s=-1$, $0$, and $1$ are 
significantly different. Obviously, the solvent quality has a noticeable 
quantitative influence on the adsorption behavior.

If $s$ is negative, the monomer-monomer interaction is repulsive, and the
polymer avoids forming nearest-neighbor contacts. This mimics the effect 
of a good solvent. In the limit $s\to -\infty$,
the system is represented by what we may call a ``super-self-avoiding walk'' 
(SSAW) model, where 
the contacts
between nearest neighbors are forbidden. This effectively increases the 
excluded volume. The adsorption
temperature of this system is expected to be smaller than for $s=0$. To 
our knowledge, this case has not yet been studied and there are no results to 
compare with. However, as our results suggest, the corresponding critical 
adsorption temperature of this intrinsically nonenergetic SSAW should be
$T_a\lesssim 3.31$.  
\begin{figure}
\centerline{\includegraphics[width=8.5cm]{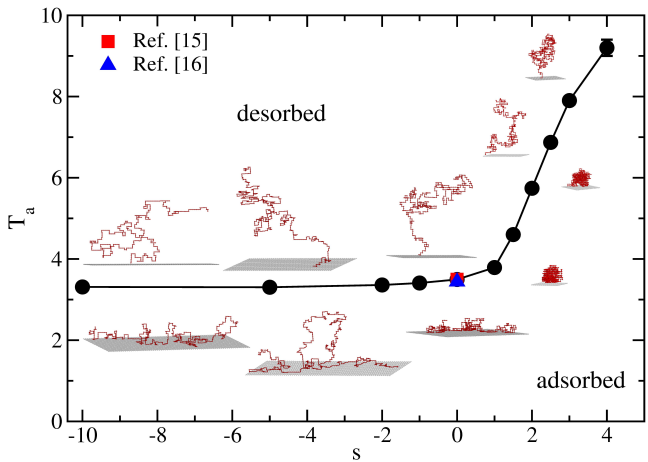}}
\caption{\label{fig:Tcs}(Color online) Critical temperature $T_a$ as a 
function of $s$ for the
adsorption-desorption transition. Results for $s=0$ from 
Refs.~\cite{binder,luo} 
are also shown for comparison. Note that 
for most data points, the error is smaller than the symbol size. Inset
conformations of the 503mer are representative for the respective regions
of parameter space. The qualitative differences indicate additional
transitions inside the adsorbed/desorbed polymer phases. For $s<0$,
super-self-avoiding conformations are dominant.}
\end{figure}

Relaxing this constraint by increasing the value of $s$ effectively increases 
the conformational entropy at a given energy in the phase of adsorbed 
conformations more than in the desorbed phase. In consequence, the slope of the 
microcanonical entropy (or the density of states) becomes smaller near the 
transition point, which, in turn, results in a larger adsorption temperature. 
The phase diagram plotted in Fig.~\ref{fig:Tcs} shows exactly this behavior for 
the adsorption temperature.
Results for $s=0$ from Refs.~\cite{luo,binder}, also included in this 
figure, fit very well into the extended picture of polymer adsorption we 
present here.

\begin{figure}
\centerline{\includegraphics[width=8.5cm]{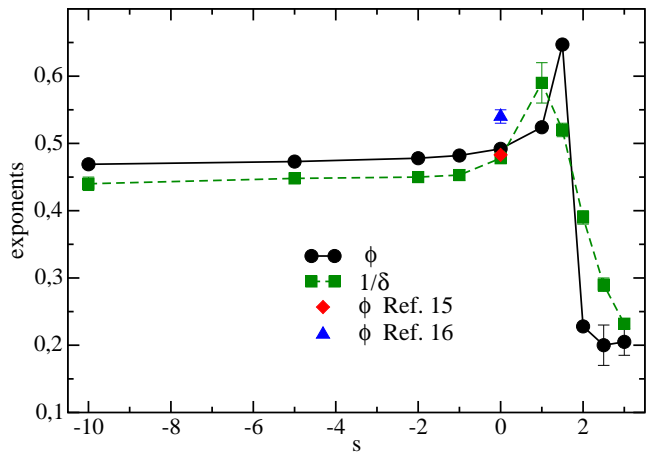}}
\caption{\label{fig:phinu}(Color online) Critical exponents $\phi$ and 
$1/\delta$ as a function of the
solvent parameter $s$. Results for $s=0$ from references~\cite{binder,luo} 
are also shown for comparison.}
\end{figure}

For all $s$ values, the adsorption transition is a second-order 
phase transition. Therefore, we are going to discuss in the following the $s$ 
dependence of the critical exponents in the entire range of the solvent 
parameter.
Figure~\ref{fig:phinu} depicts the behavior of the exponents $\phi$ and
$1/\delta$ if 
$s$ is changed. We find 
that their values vary along the second-order transition line, meaning
that 
this
transition seems to be nonuniversal. Moreover, both exponents exhibit a
peak
near $s\sim 1.5$. This can be an indication of the presence of a
multicritical
point in this region~\cite{lookman1,whit1,singh1,prellberg2}. In fact,
various additional crossovers between 
different adsorbed phases in the high-$s$ regime are expected. Analyses for
a 
finite system~\cite{mb1} show a complex structure of adsorbed compact
phases 
in this regime, but simulations of sufficiently large systems which would
allow for a thorough finite-size scaling analysis are extremely
challenging. Therefore, the discussion of the nature of
separate tricritical points or a single tetracritical point with
coil-globule transition lines extending into the desorbed and the adsorbed
phases and the crystallization behavior near the adsorption line is future
work.

In all fits of the correction-to-scaling exponent
$\omega$, we have not noted any significant dependence on 
the parameter $s$, in contrast to $\phi$ and
$1/\delta$. Furthermore, the fits are not sensitive to variations of
$\omega$. Thus, the fits of all other quantities were performed with 
the value $\omega=0.5(1)$.

In this paper, we have systematically studied critical properties of 
the adsorption transition of polymers under all solvent conditions, which was 
made possible by generalized-ensemble chain-growth simulations of a 
coarse-grained lattice model. By using finite-size-scaling theory
and properly taking into account the corrections to scaling, we have 
determined 
the critical exponents and critical temperature under various solvent 
conditions. 
A major result is the construction of the 
phase diagram in the continuous spaces of temperature and the 
parameter $s$ that 
quantifies the solvent quality. Comparison with previous results for the 
singular case of $s=0$ shows good agreement, but also the 
necessity of introducing  
an additional scaling relation and including corrections to scaling.

The structure of the phase diagram and the dependence of the critical exponents 
on the solvent parameter suggest that the
critical line does not seem to be universal under general solvent 
conditions. Moreover, the exponents exhibit a peak near $s$ values,  
where the compactness of the polymer 
conformations changes, indicating 
the 
existence of possible multicritical points of coil-globule and freezing 
transitions in the desorbed and adsorbed regimes intersecting the adsorption 
transition line.
The rather strong variation of the critical exponents, as well as the
corresponding critical temperature near this region, can be the cause for
the
difficulty encountered in quantifying the criticality of the model, even
for $s=0$.
Naturally, additional simulations in the ordered adsorbed region might be 
helpful for precisely determining the behavior of the transition lines
close to
the multicritical point, which is a separate study worth in its own right.

This work has been supported partially by CNPq (Conselho Nacional de
Desenvolvimento Cient\'ifico e Tecnol\'ogico, Brazil) under Grant No.\
402091/2012-4 and by the NSF under Grant No.\ DMR-1463241. PHLM also 
acknowledges support by FAPEMAT (Funda\c c\~ao de Amparo \`a Pesquisa do Estado 
de 
Mato Grosso) under Grant No.\ 219575/2015.
%

%
\end{document}